# Principle Component Analysis for Classification of the Quality of Aromatic Rice

Etika Kartikadarma[1*], Sari Wijayanti[2], Sari Ayu Wulandari[3], Fauzi Adi Rafrastara[1]

[1] *Faculty of Computer Science, Universitas Dian Nuswantoro, 50131, Indonesia*

[2] *Information Engineering Department, STMIK Jenderal A Yani, Yogyakarta, 552851, Indonesia*

[3]*Faculty of Engineering, Universitas Dian Nuswantoro, 50131, Indonesia*

[*]*email: etika.kartikadarma@dsn.dinus.ac.id*

***Abstract*—This research introduces an instrument for performing quality control on aromatic rice by utilizing feature extraction of Principle Component Analysis (PCA) method. Our proposed system (DNose v0.2) uses the principle of electronic nose or *enose*. *Enose* is a detector instrument that work based on classification of the smell, like function of human nose. It has to be trained first for recognizing the smell before work in classification process.**

**The aim of this research is to build an enose system for quality control instrument, especially on aromatic rice. The advantage of this system is easy to operate and not damaging the object of research. In this experiment, ATMega 328 and 6 gas sensors are involved in the electronic module and PCA method is used for classification process.**

## I. INTRODUCTION

One of the natural wealth of Indonesia is the diversity in tropical plant, and aromatic rice is a part of it, especially from group of rice. Aromatic rice is different to the common one in term of smell and quality [1]. It has a musty smell because of sugar fermentation [2]. The experiment for such kind of rice actually can be conducted in laboratories, but it takes a long time and costly.

A new hope as an alternative way to test the aromatic rice quickly and accurately has landed and it is available in the body of electronic nose (hereinafter referred as *enose*) [3]. The application of *enose* is so wide nowadays, including medical field [4]. *Enose* even can bring the benefit on the daily activity, like to monitor the ripeness level of tomatoes based on its aroma [5]. By looking at the potentiality, enose is proper to be developed more and more and to be implemented, especially in Indonesia.

## II. SYSTEM DESCRIPTION

Dinus Nose (DNose) v0.2 is a new version of enose that developed in UDINUS (Universitas Dian Nuswantoro, Indonesia), with some improvements on pattern recognition engine. Our previous system, DNose v0.1 [6], could successfully recognize 80% the smell of tofu.

This DNose v0.2 is composed of sensor array system, electronic data acquisition system, and clustering system using PCA method. In this research, DNose v0.2 is developed with the aim to determine and classify the rice quality on food industry.

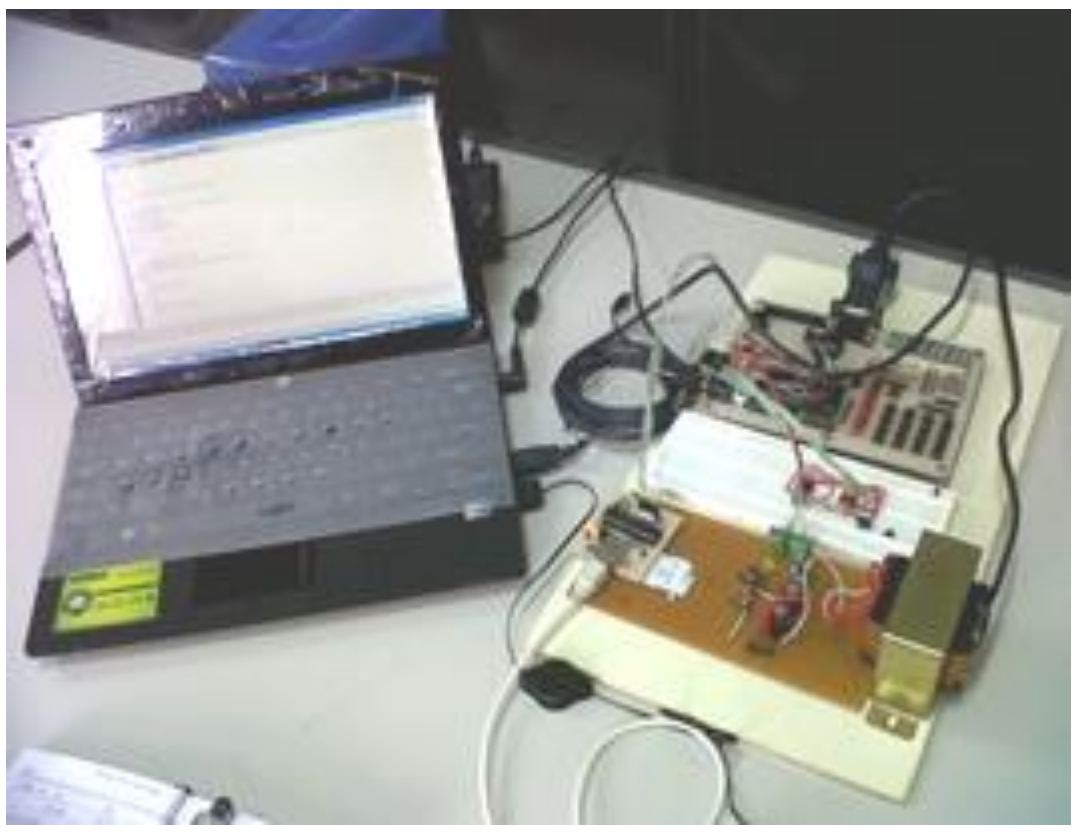

FIGURE 1. DNOSE V0.2

### A. Array Sensor System

This enose system consists of 6 gas sensors: TGS825, TGS826, TGS822, TGS813, TGS2620 and TGS2611. Those sensors are metal oxide sensor and very useful for olfactory system. Figure 2 describes the characteristic respond of sensor array when tested with aroma. Initially the aroma was sprayed into chamber sampler, then absorbed and thrown away slowly.

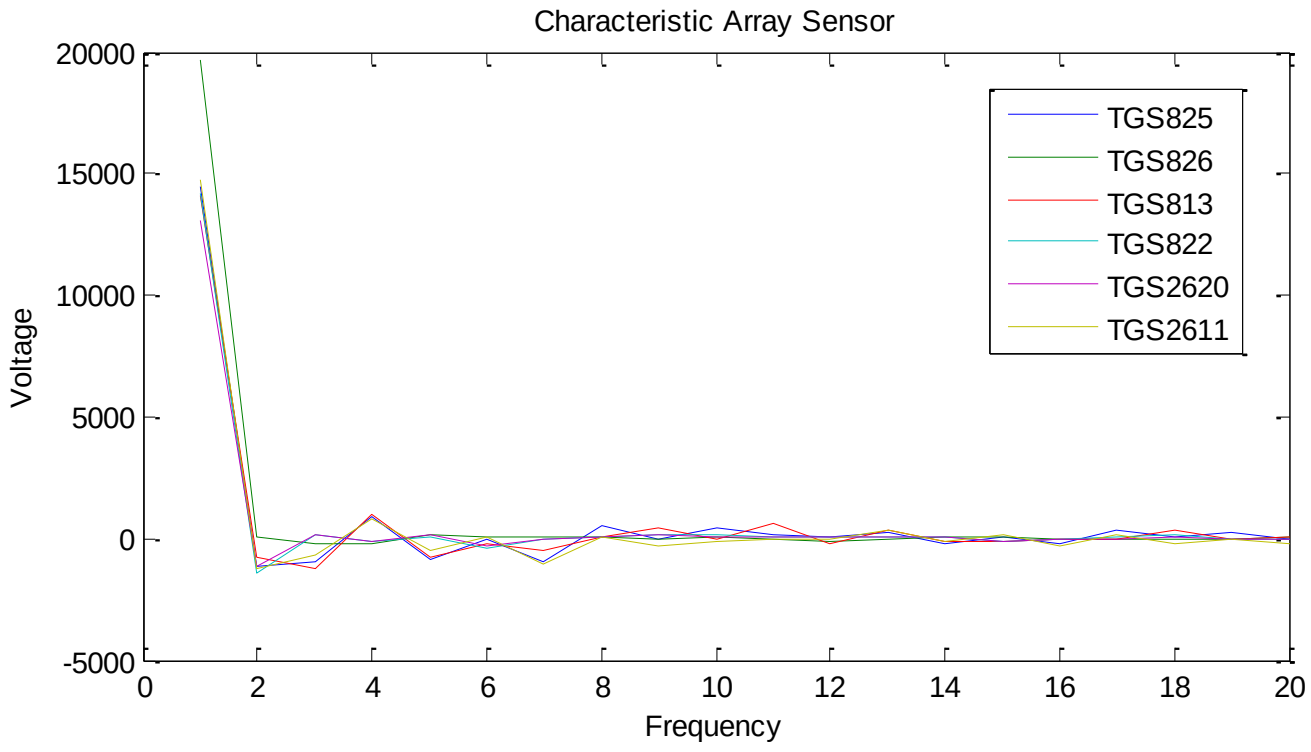

FIGURE 2. CHARACTERISTIC OF SENSOR ARRAY

### B. Clustering System

PCA is a method to extract and reduce initial pattern vector numbers into fewer pattern vectors, called principle component. In MATLAB, PCA has some functions that can be used to improve our system performance. The steps of PCA are as follows:

- Define the template vector

$$A = \begin{bmatrix} A_{11} & A_{12} & A_{13} & A_{1n} \\ A_{21} & A_{22} & A_{23} & A_{2n} \\ .... & .... & .... & .... \\ .... & .... & .... & .... \\ A_{m1} & A_{m2} & A_{m3} & A_{mn} \end{bmatrix} \quad ..........................(1)$$

- Find the reduction vector and variance matrix. The value of variance matrix represents eigen value.

$$\Sigma = \begin{bmatrix} \operatorname{var}(A_1) & \operatorname{cov}(A_1,A_2) & \operatorname{cov}(A_1,A_3) & \operatorname{cov}(A_1,A_4) \\ \operatorname{cov}(A_2,A_1) & \operatorname{var}(A_2) & \operatorname{cov}(A_2,A_3) & \operatorname{cov}(A_2,A_4) \\ \operatorname{cov}(A_3,A_1) & \operatorname{cov}(A_3,A_2) & \operatorname{var}(A_3) & \operatorname{cov}(A_3,A_4) \\ \operatorname{cov}(A_4,A_1) & \operatorname{cov}(A_4,A_2) & \operatorname{cov}(A_4,A_3) & \operatorname{var}(A_4) \end{bmatrix}$$

(2)

- The result of theorem 2 will be used to find the covariance matrix. The value of covariance matrix represents the eigen value.
- To create Matrix L we can check the diagonal value of theorem 2 (Matrix A). That diagonal matrix represents the eigen value.

$$L = \begin{bmatrix} \lambda_1 & 0 & 0 & 0 & 0 \\ 0 & \lambda_2 & 0 & 0 & 0 \\ 0 & 0 & .. & 0 & 0 \\ 0 & 0 & 0 & ... & 0 \\ 0 & 0 & 0 & 0 & \lambda_n \end{bmatrix} \quad ........................(3)$$

- Where $\lambda_1 > \lambda_2 > ...... > \lambda_n$.
- Find the set of eigen vector $\{u_1, u_2, .........\, .u_n\}$ from matrix S where $v_i$ is orthonormal eigen vector and consistent to the eigen value $\lambda_i$.
- Principal Component is obtained from the equation below [7].

$$PC_i = u_i.A' \quad ................................(4)$$

### C. Normalization

Gas sensing system which is also known as system for aroma detection and identification, become an important research for examining the quality of food material. An enose with VOC (Volatile Organic Compound) system can works very well in a special place with certain air temperature, signal booster, and gas sensor. Figaro gas sensor is also using VOC system. However, there is a main problem on VOC system that it cannot identified well on the patterns which almost alike. It has low sensitivity on sensing system [7]. On this case, the success rate of PCA with VOC is considered low since there is no algorithm to perform the learning process from new aroma.

Actually, the process of pattern normalization can affect to the loss of information on concentration level, but somehow, some algorithms like NN, GA, and Cluster Analysis are using the normalized response.

The most common normalization methods for enose research are amplitude and frequency normalization. The result of amplitude normalization is linier amplitude where the highest value of normalized amplitude is 1. Amplitude normalization method keeps the form of the signal and it only moves on positive axis of Cartesian coordinate y.

One of the popular amplitude normalization methods is Power Average. This method is begun by finding the highest score of pattern vector. It then performs division between pattern vector and the maximum score of pattern vector. After completing this step, then we find the average of normalized pattern vector for each class. Finally, the obtained average score is taken as template pattern that going to be compared with experiment data.

On frequency normalization, the result is in the form of polarized amplitude, and it is moving on the positive and negative axis from Cartesian coordinate y. Frequency normalization method that commonly used is fft (fast fourier transform). This method is begun by finding fft score of each pattern vector. Next, the normalized pattern vector is calculated to find the average score for each cluster. Finally, this average score is compared with experiment data, as a pattern template.

(15)

## III. EXPERIMENT

The aim of this research is to propose a DNose v0.2 based on PCA method for recognizing the smell of aromatic rice. The steps are as follows:

1. Measuring odorant, weight: 1 ons.
2. Collecting censor data for 1 minute in a room with $27^0$C temperature, and volume of odorant chamber is 1573 $cm^3$.
3. Perform sampling odorant data to find the Analog Digital Converting (ADC) value from Microcontroller, fs = 3Hz.
4. Performing data normalization by using 2 methods: power average and fft (fast fourier transform) method.
5. Applying PCA clustering with the results of those 2 normalization method.
6. Repeating the experiment up to 3 times for each method.

## IV. RESULT AND DISCUSSION

### A. Experiment of Power Average Normalization

Power average method is done by processing voltage data that yielded by those six censors, which initially have 60 samples then reduced to 20 samples, with sampling frequency fs = 3 Hz. The average of FFT normalization is listed in Table 1, whereas the graph for normalization result of censor 1 to 6 is illustrated in Figure 3.

TABEL I

RESULT OF POWER AVERAGE NORMALIZATION

| Quality | Censor 1 | Censor2 | Censor3 | Censor4 | Censor5 | Censor6 |
|---|---|---|---|---|---|---|
| KW1 | 2.14E+08 | 3.89E+08 | 2.05E+08 | 2.03E+08 | 1.73E+08 | 2.23E+08 |
| KW2 | 56845310 | 1.7E+08 | 1.95E+08 | 1.39E+08 | 1.07E+08 | 2.68E+15 |
| KW3 | 84248108 | 3.07E+08 | 93200111 | 76526991 | 69535669 | 3.27E+08 |

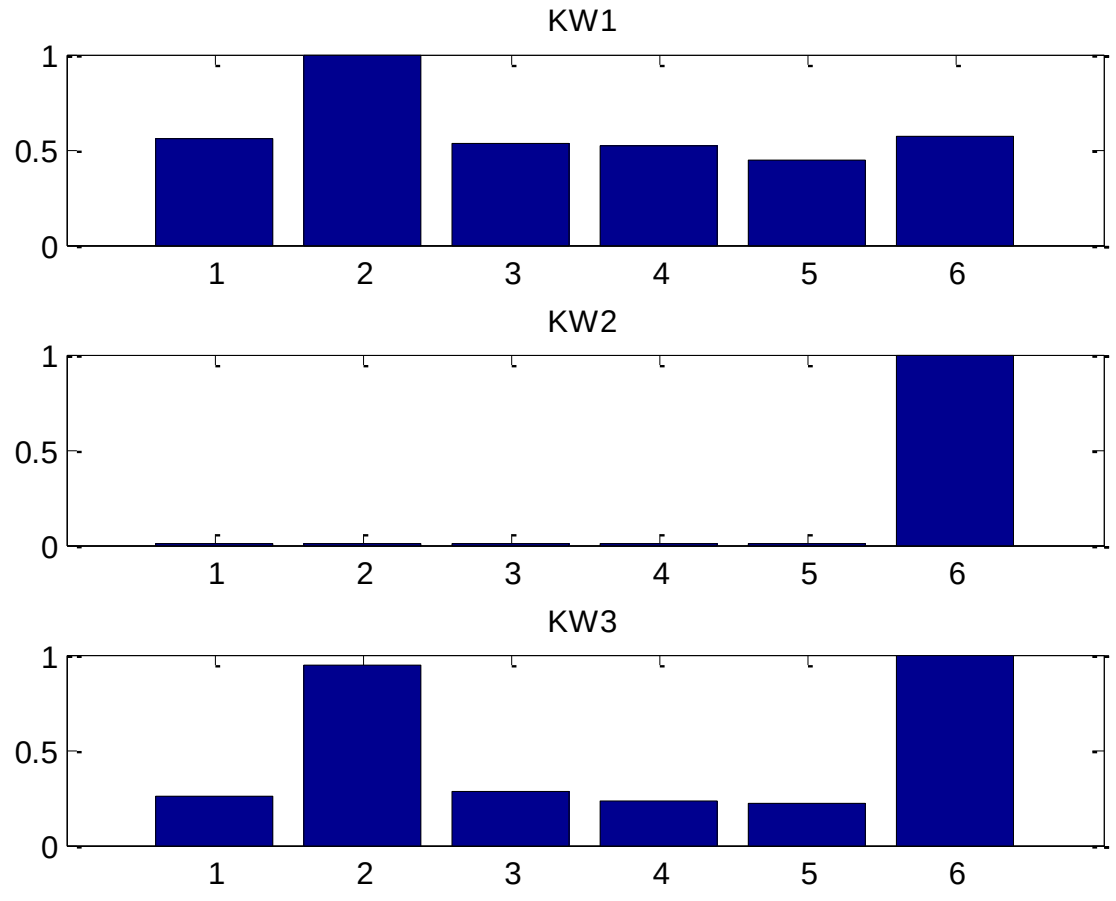

FIGURE 3. THE RESULT OF POWER AVERAGE NORMALIZATION

The gathered result of power average normalization is clustered by using PCA, as illustrated in Figure 4, 5, 6, and 7.

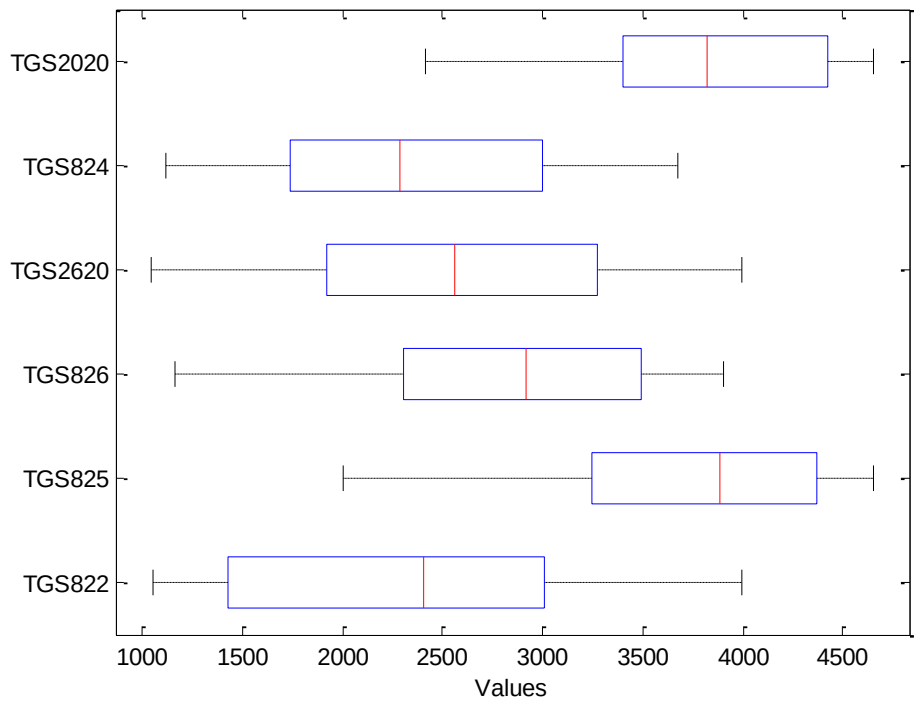

FIGURE 4. BLOCK PLOT OF POWER AVERAGE NORMALIZATION ON PCA

TABEL 2
DEVIATION STANDARD ON PCA POWER AVERAGE NORMALIZATION

| Censor 1 | Censor 2 | Censor 3 | Censor 4 | Censor 5 | Censor 6 |
|---|---|---|---|---|---|
| 869.4426 | 710.0997 | 675.6139 | 646.1813 | 580.0782 | 626.4368 |

Table 2 shows that deviation standard of PCA with power average normalization is still high, with the highest deviation is on censor 1 (869.4426) and the lowest deviation is on censor 5 (580.0782). Those data can be a reference to extract some censors that have high deviation standard as it can obstruct the pattern recognition process.

The clustering result of power average normalization shows that the clusters are still distributed and the distance of each cluster is still close, as illustrated in Figure 6.

The pareto chart illustrates the plots of variable percentage which explained on each Principle Component (PC). Figure 6 shows the clear differences between first and second component. However, the variance differences between those two components is still less than 50%, so it is required more components to be involved. According to Figure 6, the first two components represent 2/3 of total variabilities on the template.

Here, we can use FFT to reduce the dimension so that we can visualize the data.

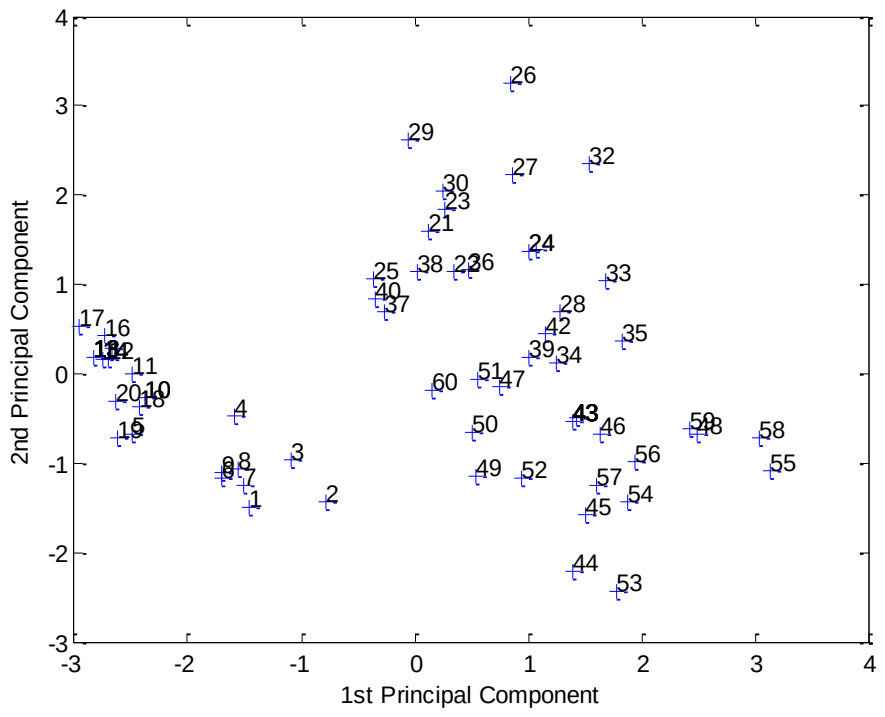

FIGURE 5. THE FIRST AND SECOND PRINCIPAL COMPONENTS WITH POWER AVERAGE NORMALIZATION METHOD

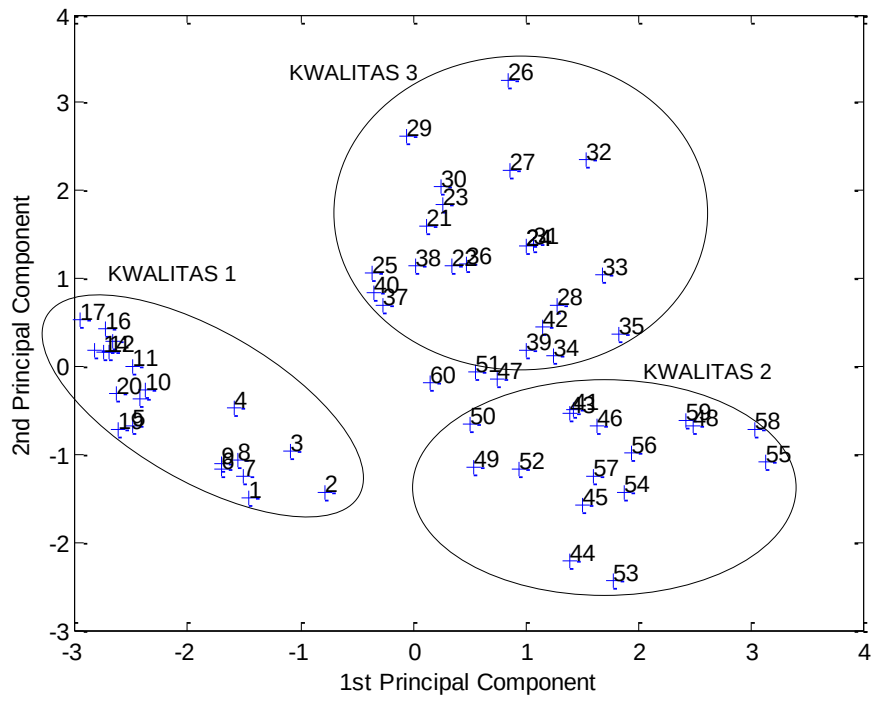

FIGURE 6. THE FIRST AND SECOND PRINCIPAL COMPONENTS CLUSTER DESCRIPTION WITH POWER AVERAGE NORMALIZATION

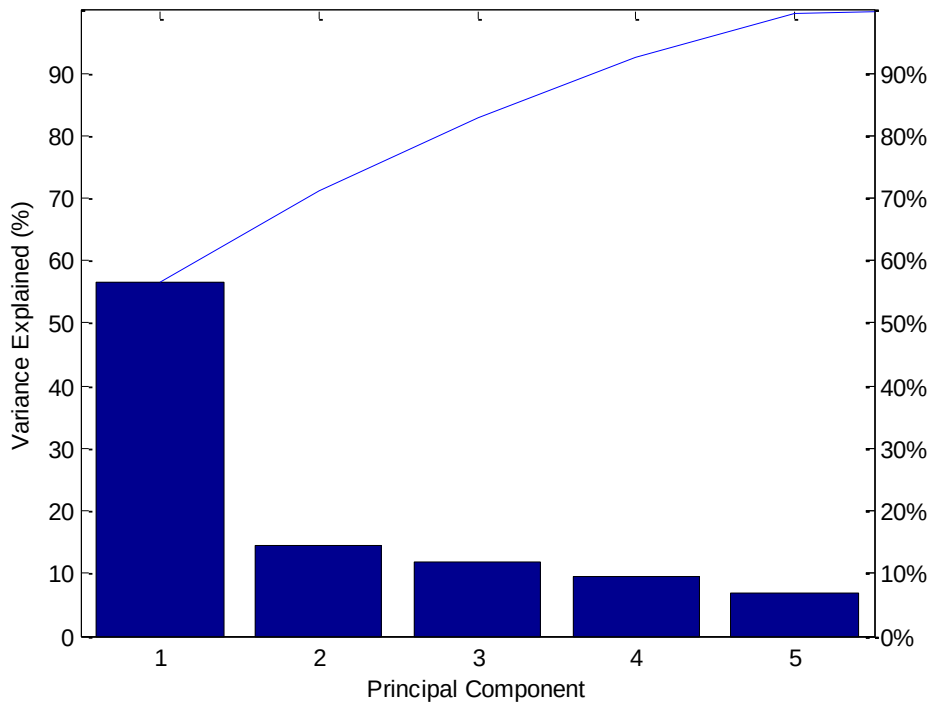

FIGURE 7. PARETO DIAGRAM OF EACH PRINCIPAL COMPONENT WITH POWER AVERAGE NORMALIZATION

## B. *Experiment of FFT Normalization*

Power average method is applied by processing the voltage data that obtained from those six censors. Initially it has 60 samples and then reduced to 20 samples, with sampling frequency fs = 3 Hz. The average of FFT normalization is listed in Table 3, whereas the graph for normalization result of censor 1 to 6 is illustrated in Figure 7.

TABEL 3

THE RESULT OF FFT NORMALIZATION

| Quality | Censor 1 | Censor2 | Censor3 | Censor4 | Censor5 | Censor6 |
|---|---|---|---|---|---|---|
| KW1 | 4.2E+09 | 7.77E+09 | 4.02E+09 | 3.97E+09 | 3.4E+09 | 4.36E+09 |
| KW2 | 1.02E+09 | 3.29E+09 | 3.82E+09 | 2.76E+09 | 2.05E+09 | 1.91E+19 |
| KW3 | 1.54E+09 | 6.13E+09 | 1.79E+09 | 1.44E+09 | 1.33E+09 | 6.41E+09 |

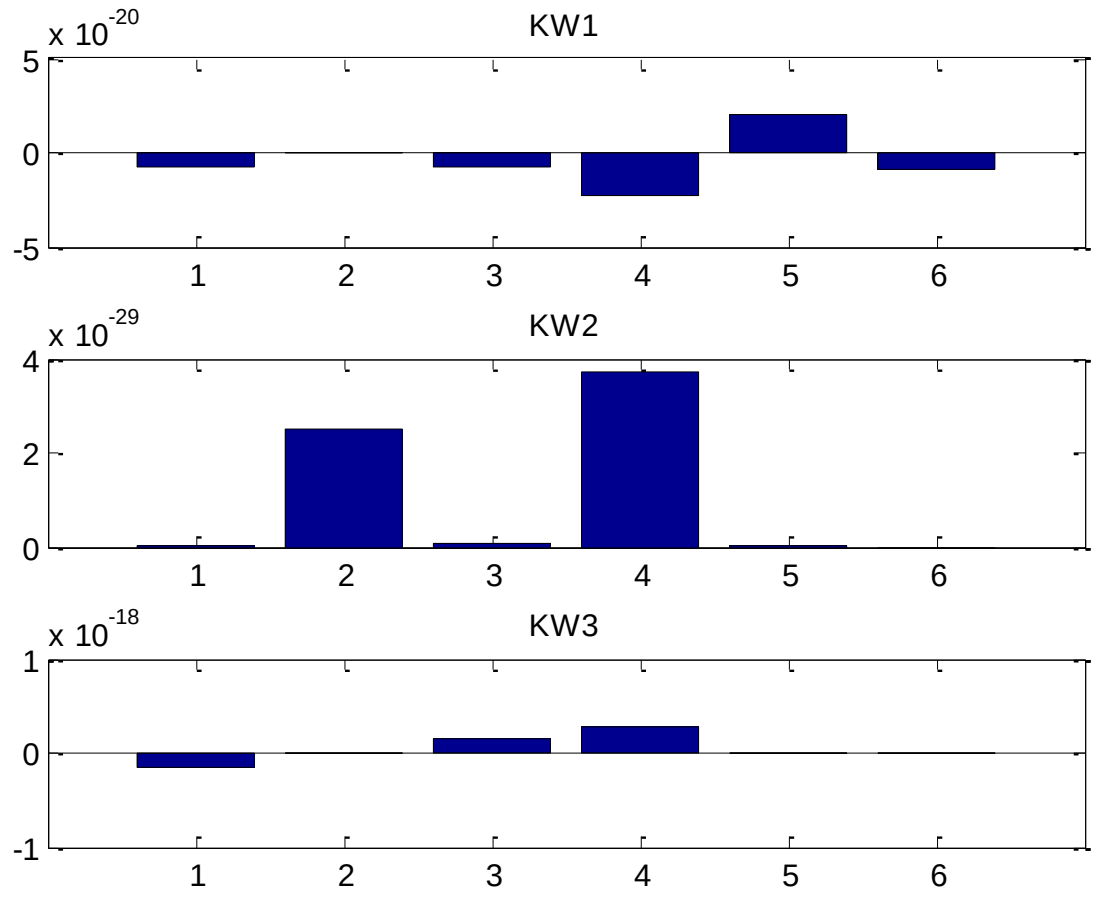

FIGURE 8. RESULT OF FFT NORMALIZATION

On this step, the result of FFT normalization is clustered by using PCA, as shown in figure 8, 9, and 10.

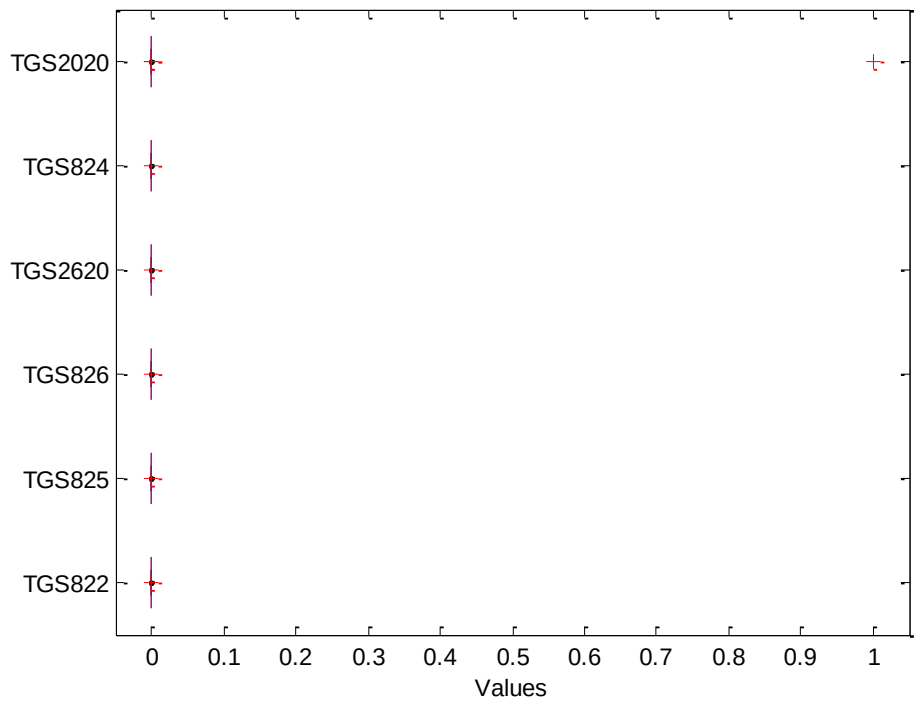

FIGURE 9. BLOCK PLOT OF FFT NORMALIZATION ON PCA

TABLE4
DEVIATION STANDARD ON PCA WITH FFT NORMALIZATION

| Censor 1 | Censor 2 | Censor 3 | Censor 4 | Censor5 | Censor6 |
|---|---|---|---|---|---|
| 0 | 0 | 0 | 0 | 0 | 0.1291 |

Table 4 explains the deviation standard of PCA (FFT normalization) which has slight better score, with the highest deviation is only on censor 6 (0.1291) and the lowest deviation is on censor 1 to 5 (0). Those data can be a reference to remove the censor that has high deviation standard. This is because a censor with high deviation standard can obstruct the pattern recognition process. So, in this case, we remove censor 6.

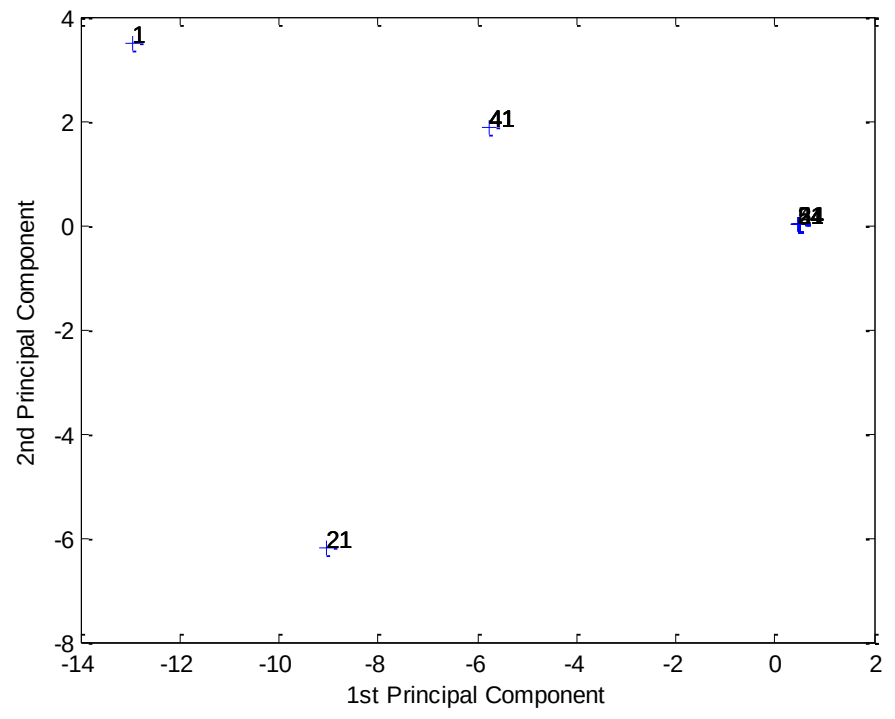

FIGURE 10. THE FIRST AND SECOND PRINCIPAL COMPONENTS

The clustering result of power average normalization shows that the clusters have been well concentrated, and the distance among each cluster is far enough. There is a new cluster created by merging between cluster 2 and 3. Table 5 shows the list of clusters involved.

TABLE5
PATTERN VECTOR DISTRIBUTION OF PCA WITH FFT NORMALIZATION

| Quality | Rows |
|---|---|
| KW1 | 1,2,3,4,5,6,7,8,9,10,11,12,13,14,15,16,17,18,19,20 |
| KW2 | 21,22,23,25,26,27,28,29,31,33,34,35,36,37,39,40 |
| KW3 | 41,43,44,45,46,48,49,50,51,52,53,54,56,57,58,59 |
| New Cluster | 24,30,32,38,42,47,55,60 |

According to Table 5, Q1 (Quality 1) has distribution 0%, whereas the distribution of both Q2 and Q3 are 20%.

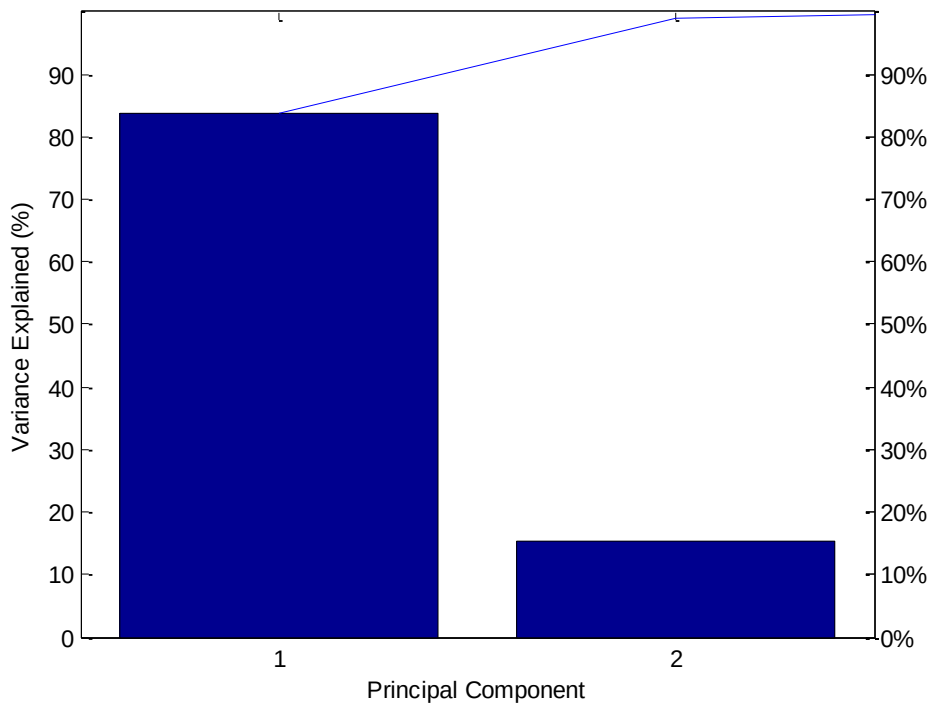

FIGURE 11. PARETO DIAGRAM OF EACH PRINCIPAL COMPONENT

By applying the result of PCA into pattern recognition algorithm, the smaller distribution score on PCA system means the pattern can be recognized easier.

The above pareto chart illustrates the plots of variable percentage which explained on each Principle Component (PC). Figure 10 shows the big differences between first and second component. However, the variance differences between those two components is still less than 84%, so it is required more components to be involved. According to Figure 6, the first two components represent 2/3 of total variabilities on the template. Here, we can use FFT to reduce the dimension so that we can visualize the data.

## V. CONCLUSION

Based on the experiment and gathered result, we conclude that FFT normalization method is better than power average one. It is proved by the score of *variability explained* on FFT is higher than the score of *variability explained* on power average normalization method. However, by using FFT, the result of clustering process is still less than 0.1%. Therefore, the further research is needed to find the better normalization method that can produce significant improvement of the percentage of variability explained.

## ACKNOWLEDGEMENT

We would like to express our gratitude to Dr. Eng. Kuwat Triana, M.Si, he has been our mentor for this research: eNose for detection of quality of aromatic rice.